\documentstyle[aps,psfig]{revtex}
\begin{document}
\draft
\title{Vector meson masses in hot nuclear matter: the effect of
quantum corrections} 

\author{Amruta Mishra\footnote[1]
{email: mishra@th.physik.uni-frankfurt.de}$^\dagger$, 
Jitendra C. Parikh $^\ddagger$, Walter Greiner$^\dagger$}
\address {$^\dagger$Institut f\"ur Theoretische Physik, 
J.W. Goethe Universit\"at, Robert Mayer-Stra{\ss}e 10,\\
Postfach 11 19 32, D-60054 Frankfurt/Main, Germany}
\address{$^\ddagger$ Theory Division, Physical Research Laboratory,
Ahmedabad - 380 009, India}

\maketitle
\begin{abstract}
The medium modification of vector meson masses is studied taking into
account the quantum correction effects for the hot and dense hadronic
matter. In the framework of Quantum Hadrodynamics, the quantum corrections
from the baryon and scalar meson sectors were earlier computed using a 
nonperturbative variational approach through a realignment of the
ground state with baryon-antibaryon and sigma meson condensates.
The effect of such corrections  was seen to lead to a softer
equation of state giving rise to a lower value for the compressibility
and, an increase in the in-medium baryonic masses  than would be
reached when such quantum effects are not taken into account.
These quantum corrections arising from the scalar meson sector
result in an increase in the masses of the vector mesons
in the hot and dense matter, as compared to the situation when
only the vacuum polarisation effects from the baryonic sector
are taken into account.
\end{abstract}

\def\bfm#1{\mbox{\boldmath $#1$}}

\section{Introduction}
The effective masses of the vector mesons ($\omega$,$\rho$ and
$\phi$) in the hot and dense matter have attracted a lot of interest 
in the recent past, both experimentally \cite {helios,ceres} and 
theoretically \cite{brown,hat,jin,samir,weise}. Brown and Rho came 
with the hypothesis that the vector meson masses drop in the medium
 according to a simple (BR) scaling law \cite{brown}, given as 
$m_V^*/m_V=f_\pi^*/f_\pi$, where $f_\pi$ is the pion decay constant 
and asterix refers to in-medium quantities. 
There have been approaches based on QCD sum rules \cite {hat,jin}
which confirm such a scaling hypothesis \cite {hat} with a saturation
scheme that leads to a delta function at the vector meson pole and
a continuum for higher energies for the hadronic spectral function.
It is however seen that such a simple saturation scheme for finite
densities does not work and a more realistic description for the
hadronic spectral function is called for \cite{samir}. Using an
effective Lagrangian model to calculate the hadronic spectral 
function, it is seen that such a universal scaling law is not 
suggested for in-medium vector meson masses \cite{weise}.

In the conventional hadronic models \cite {serot,chin}, on the other hand, 
the masses of the vector mesons stay constant or increase slightly,
in the mean field approximation, i.e., when the polarization
from the Dirac sea is neglected \cite{chin}. With the inclusion 
of quantum corrections from the baryonic sector, however, one observes
a drop in the vector meson masses in the medium in the Walecka model
\cite{hatsuda,hatsuda1,jeans}. 
This medium modification of the vector meson masses plays an important role
in  the enhanced dilepton yield \cite{wambach} for masses below the 
$\rho$ resonance in the heavy ion collision experiments \cite{helios,ceres}. 
It has been emphasized recently that the Dirac sea contribution 
(taken into account through summing over baryonic tadpole diagrams 
in the relativistic Hartree approximation (RHA)) dominates over 
the fermi sea contribution and $m_\omega^*/m_\omega <1$ is caused 
by a large dressing of $\bar N N$ cloud in the medium \cite{jeans}. 
It was earlier demonstrated in a nonperturbative formalism 
that a realignment of the ground  state with baryon-antibaryon 
condensates was equivalent to the relativistic Hartree approximation
\cite {mishra}.
The ground state for the nuclear matter was extended
to include sigma condensates to take into account the quantum
correction effects from the scalar meson sector. Such a formalism
includes multiloop effects and is self consistent. The methodology 
was then generalized to consider hot nuclear matter \cite{hotnm}
as well as to the situation of hyperon-rich dense matter \cite{shm}.
In the present work, we study the effect of such quantum corrections
on the in-medium vector meson masses.

We organize the paper as follows. We first briefly recapitulate the
nonperturbative framework used for studying the quantum correction
effects in hot nuclear matter in section II. The medium modification
of the $\omega$ and $\rho$ meson masses are considered in section III.
We also examine the effect of quantum corrections on the in-medium 
vector meson masses in dense hyperon-rich matter in section IV, 
where it is observed that the medium modification to the strange vector 
meson ($\phi$) is small compared to that of the vector $\omega$ meson. This 
is because, in the hyperonic matter with quantum effects taken into
account \cite{shm}, the masses of the hyperons remain 
rather insensitive to density, 
unlike the in-medium  nucleon masses which change appreciably
with density, and further, because the $\phi$-mesons do not couple to the
nucleons \cite {hatsuda1}. On the other hand,  $\omega$ meson mass 
gets contribution also from the nucleonic sector. 
In section V, we summarize the results of the present work and discuss 
possible outlook.

\section {Quantum Vacuum in hot nuclear matter}

We briefly recapitulate here the vacuum polarisation effects arising 
from the nucleon and scalar meson fields in hot nuclear matter in a 
nonperturbative framework \cite{hotnm}. The method of thermofield 
dynamics (TFD) is used here to study the ``ground state" (the state 
with minimum thermodynamic potential) at finite temperature and density 
within the Walecka model. The temperature and density dependent 
baryon and sigma masses are also calculated in a self-consistent
manner in the formalism. The ansatz functions involved in such an 
approach are determined through minimisation of the thermodynamic 
potential. 

The Lagrangian density in the Walecka model is given as
\begin{eqnarray}
{\cal L}&=&\bar \psi (i\gamma^\mu \partial_\mu
-M-g_\sigma \sigma-g_\omega\gamma^\mu \omega_\mu)\psi
+\frac{1}{2}\partial^\mu\sigma
\partial_\mu\sigma-\frac{1}{2} m_\sigma ^2 \sigma^2
-\lambda \sigma^4
\nonumber\\
&+&\frac{1}{2} m_\omega^2 \omega^\mu \omega_\mu
-\frac{1}{4}(\partial_\mu \omega_\nu -\partial_\nu \omega_\mu)
(\partial^\mu \omega^\nu -\partial^\nu \omega^\mu).
\end{eqnarray}
In the above, $\psi$, $\sigma$, and $\omega_\mu$ are the fields
for the nucleon, $\sigma$, and $\omega$ mesons
with masses M, $m_\sigma$, and $m_\omega$ respectively. 
The quartic coupling term in $\sigma$ is necessary for the 
sigma condensates to exist, through a vacuum realignment \cite{mishra}. 
We retain the quantum nature of both the nucleon and the scalar meson 
fields, whereas, the vector $\omega$-- meson is treated as a 
classical field, using the mean field approximation for $\omega$--meson,
given as $\langle \omega^\mu \rangle=\delta_{\mu 0} \omega_0$. The reason 
is that without any quartic or any other higher order term for the 
$\omega$-meson, the quantum effects generated due to $\omega$-meson 
through the present variational ansatz turns out to be zero.

The Hamiltonian density can then be written as
\begin{equation}
{\cal H}={\cal H}_N+{\cal H}_\sigma+{\cal H}_\omega,
\end{equation}
with
\begin{mathletters}
\begin{equation}
{\cal H}_N=\psi ^\dagger(-i {\bf \alpha} \cdot {\bfm \bigtriangledown}
+\beta M)\psi + g_\sigma\sigma \bar \psi\psi,
\end{equation}
\begin{equation}
{\cal H}_\sigma=
\frac {1}{2} {\dot \sigma}^2+
\frac{1}{2} \sigma (-{\bfm \bigtriangledown}^2)
\sigma+\frac{1}{2} m_\sigma ^2 \sigma^2+\lambda \sigma^4,
\label{lwsg}
\end{equation}
\begin{equation}
{\cal H}_\omega= g_\omega \omega_0 \psi
^\dagger \psi
 -\frac{1}{2} m_\omega ^2 \omega_0^2.
\end{equation}
\end{mathletters}
We may now write down the field expansion for the nucleon field
$\psi$ at time $t=0$ as \cite{mishra}
\begin{equation}
\psi(\bfm x)=\frac {1}{(2\pi)^{3/2}}\int \left[U_r(\bfm k)c_{Ir}(\bfm k)
+V_s(-\bfm k)\tilde c_{Is}(-\bfm k)\right] e^{i\bfm k\cdot \bfm x} d\bfm k,
\end{equation}
with $c_{Ir}$ and $\tilde c_{Is}$ as the baryon  annihilation 
and antibaryon creation operators with spins $r$ and $s$ respectively,
and $U$ and $V$ are the spinors associated with the particles and
antiparticles respectively \cite{mishra}.
Similarly, we may expand the field operator of the scalar field $\sigma$ 
in terms of the creation and annihilation operators, at time $t=0$ as
\begin{equation}
\sigma (\bfm x,0)={1\over{(2 \pi)^{3/2}} }\int{{d\bfm k\over{\sqrt
{2 \omega (\bfm k)}}}\left(a(\bfm k)+
a^\dagger(-\bfm k)\right)e^{i\bfm k\cdot\bfm x}}.
\label{expan}
\end{equation}
In the above, $\omega (\bfm k)=\sqrt{\bfm k^2+m_\sigma^2}$.
The perturbative vacuum is defined corresponding to this basis through 
$a\mid vac\rangle=0=c_{Ir}\mid vac \rangle
=\tilde c_{Ir}^\dagger\mid vac \rangle$. 

To include the vacuum polarisation effects for hot nuclear matter, 
we shall now consider a trial state with baryon--antibaryon and 
scalar meson condensates and then generalize the same to the
finite temperatures and densities \cite{mishra}. 
We thus explicitly take the ansatz for the trial state as 
\begin{equation}
|F\rangle =U_\sigma U_F|vac\rangle,
\label{cond}
\end{equation}
with
\begin{equation}
U_F= \exp \Big[ \int d\bfm k ~f(\bfm k)~{c_{Ir}^\dagger (\bfm k)}~
a_{rs} \tilde c _{Is} (-\bfm k)-h.c. \Big]
\end{equation}
Here $a_{rs}=u_{Ir}^\dagger(\bfm \sigma \cdot \hat k)v_{Is}$ 
and $f(\bfm k)$ is a trial function associated with baryon-antibaryon 
condensates.
For the scalar meson sector,
$ U_\sigma=U_{II}U_{I}$
where $U_{i}=\exp(B_i^\dagger~-~B_i),\,(i=I,II)$. Explicitly the
$B_{i}$ are given as
\begin{equation}
B_I^\dagger=\int {d\bfm k \sqrt{\omega (\bfm k)\over 2}
f_\sigma(\bfm k) a^\dagger(\bfm k)},\;\;\;
{B_{II}}^\dagger={1\over 2}\int d\bfm k g(\bfm k){a'}^\dagger(\bfm k)
{a'}^\dagger(-\bfm k). 
\end{equation}
In the above, $a'(\bfm k)=U_I a(\bfm k) U_I^{-1}=a(\bfm k)-
\sqrt{\frac{\omega (\bfm k)}{2}}f_\sigma(\bfm k)$ corresponds to a
shifted field operator associated with the coherent state 
\cite{mishra,amhm} 
and satisfies the usual quantum algebra. Further, to preserve 
translational invariance $f_\sigma(\bfm k)$ has to be  proportional to $\delta
(\bfm k)$ and  we take $f_\sigma(\bfm k)=\sigma _{0} (2\pi)^{3/2}
\delta (\bfm k)$. $\sigma_0$ corresponds to a
classical field of the conventional approach \cite{mishra}.
Clearly, the ansatz state is not annihilated by the operators,
$c$, $\tilde c^\dagger$ and $a$. However, one can define operators,
$d$, $\tilde d^\dagger$ and $b$, related through a Bogoliubov
transformation to these operators, which will annihilate 
the state $|F\rangle$.


We next use the method of thermofield dynamics \cite{tfd} to construct 
the ground state for nuclear matter at finite temperature. Here the 
statistical average of an operator is written as an expectation value 
with respect to a `thermal vacuum' constructed from operators defined 
on an extended Hilbert space. The `thermal vacuum' is obtained from the 
zero temperature ground state through a thermal Bogoliubov transformation.
We thus generalise the state, as given by (\ref{cond}) to finite 
temperature and density as
\cite{mishra,amhm}
\begin{equation}
|F,\beta\rangle=U_\sigma(\beta)U_F(\beta)|F\rangle.
\label{trialansatz}
\end{equation}
The temperature-dependent unitary operators $U_\sigma(\beta)$ and
$U_F(\beta)$ are given as \cite{tfd}

\begin{equation}
U_\sigma(\beta)=\exp{\Bigg({1\over 2}\int d\bfm k \theta_\sigma(\bfm k,\beta) 
b^\dagger(\bfm k) {\underline b}^\dagger(-\bfm k)-h.c.\Bigg)}. 
\end{equation}
and 
\begin{equation}
U_F(\beta) =\exp{\Bigg(
\int d\bfm k ~\bigg[\theta_-(\bfm k,\beta)~
d_{Ir}^\dagger (\bfm k)~{\underline d}_{Ir}^\dagger(-\bfm k)
+\theta_+(\bfm k,\beta)~ \tilde d_{Ir}(\bfm k)~\tilde 
{\underline d}_{Ir}(-\bfm k)\bigg]-h.c.\Bigg)}.
\end{equation}
The underlined operators are the operators corresponding to the
doubling of the Hilbert space that arise in thermofield dynamics method.
We shall determine the condensate functions $f(\bfm k)$ and $g(\bfm k)$,
and the functions  $\theta_\sigma(\bfm k,\beta)$, $\theta_-(\bfm k,\beta)$ 
and $\theta_+(\bfm k,\beta)$ 
of the thermal vacuum through minimisation of the thermodynamic potential.
The thermodynamic potential is given as
\begin{equation}
\Omega \equiv -p=\epsilon-\frac{1}{\beta}{\cal S}-\mu \rho_B,
\label{thermpot}
\end{equation}
where $\epsilon$ and ${\cal S}$ are the energy- and entropy-
densities of the thermal vacuum, and $\rho_B$ is the baryon density.
The ansatz functions used in the definition of the 
thermal vacuum are determined  through the minimisation of the 
thermodynamic potential. Then subtracting out the pure vacuum
contribution and carrying out the renormalisation procedures for the
baryonic and scalar meson sectors \cite{hotnm}, 
we obtain the expression for the energy density as
\begin{equation}
\epsilon_{ren}=\epsilon_{finite}^{(N)}
+\Delta \epsilon_{ren}+\epsilon_\omega+\Delta \epsilon_\sigma,
\end{equation}
with,
\begin{mathletters}
\begin{equation}
\epsilon_{finite}^{N}=
\gamma (2\pi)^{-3}\int d \bfm k (k^2+{M^*}^2)^{1/2} 
(\sin^2 \theta_- +\sin ^2 \theta_+) 
\end{equation}
\begin{eqnarray}
\Delta \epsilon_{ren} &=& -\frac{\gamma}{16\pi^2}
 ( {M^*}^4 \ln \Big (\frac{M^*}{M}\Big )
+M^3 (M-M^*)-\frac{7}{2} M^2 (M-M^*)^2 \nonumber\\
& + & \frac{13}{3} M (M-M^*)^3 -
 \frac{25}{12} (M-M^*)^4 )
\end{eqnarray}
\begin{equation}
\epsilon_{\omega}=
g_\omega \omega_0 \rho_B^{ren}-\frac {1}{2} m_\omega^2 \omega_0^2,
\end{equation}
\begin{eqnarray}
\Delta \epsilon_\sigma 
&=& \frac{1}{2} m_R^2 \sigma_0^2+ 3\lambda_R \sigma_0^4 
+\frac {M_\sigma^4}{64\pi^2}
\Biggl(\ln\Big(\frac{M_\sigma^2}{m_R^2}\Big)-\frac{1}{2} \Biggr)
-3\lambda_R I_f^2\nonumber\\
&-&\frac {M^4_{\sigma,0}}{64\pi^2}
\Biggl(\ln\Big(\frac{M_{\sigma,0}^2}{m_R^2}\Big)-\frac{1}{2} \Biggr)
+3\lambda_R I_{f0}^2,
\label{vph0}
\end{eqnarray}
\end{mathletters}

\noindent as the mean field result, contribution from the Dirac sea,
and contributions from the $\omega$ and $\sigma$ mesons respectively.
In the above, $\sin^2 \theta_{\mp}$ are the distribution functions for
the baryons and antibaryons given through
\begin{equation}
\sin^2 \theta _{\mp}=\frac{1}{\exp(\beta(\epsilon^*(k)\mp \mu^{*})) 
+1},
\label{distr}
\end{equation}
with $\epsilon^*(k)=(k^2+{M^*}^2)^{1/2}$ and $\mu^{*}=\mu -g_\omega
\omega_0$ as the effective energy and effective chemical
potential, where the effective nucleon mass $M^{*}=M+g_\sigma \sigma_0$.
The baryon number density after subtracting out the pure vacuum 
contribution is given as
\begin{equation}
\rho_B^{ren}=\gamma (2\pi)^{-3}\int d\bfm k 
(\sin^2 \theta_- -\sin ^2 \theta_+).
\end{equation}
In the expression for energy density arising from the scalar meson sector,
the field dependent effective sigma mass, 
$M_\sigma(\beta)$, satisfies the gap equation
 in terms of the renormalised parameters as
\begin{equation}
M_\sigma(\beta)^2=m_R^2+12\lambda_R\sigma_0^2+12\lambda_R I_f(M_\sigma(\beta)),
\label{mm2}
\end{equation}
where,
\begin{equation}
I_f(M_\sigma(\beta))=\frac{M_\sigma(\beta)^2}{16\pi^2}
\ln \Big(\frac{M_\sigma(\beta)^2}{m_R^2} \Big)+
\frac{1}{(2\pi)^3}\int d\bfm k \frac{\sinh^2\theta_\sigma(\bfm k,
\beta)}{(\bfm k^2+ M_\sigma(\beta)^2)^{1/2}},
\label{if}
\end{equation}
and,
\begin{equation}
\sinh^2 \theta _\sigma=\frac{1}{e^{\beta\omega'(\bfm k,\beta) -1}};
\;\;\;
\omega'(\bfm k,\beta)=(\bfm k^2+M_\sigma(\beta)^2)^{1/2}.
\end{equation}

In equation (\ref{vph0}), $M_{\sigma,0}$ and $I_{f0}$ are the expressions
as given by eqs. (\ref{mm2}) and (\ref{if}) with $\sigma_0=0$.
We might note here that the gap equation given by (\ref{mm2})
is identical to that obtained through resumming the daisy 
and superdaisy graphs \cite{pi} and hence includes
higher order corrections from the scalar meson field.

The thermodynamic potential, $\Omega$, given  by equation 
(\ref {thermpot}), after subtracting out the vacuum contributions,
is now finite and is given in terms of the meson fields,
$\sigma_0$ and $\omega_0$.
Extremisation  of the thermodynamic potential with respect
to the meson fields $\sigma_0$ and $\omega_0$ give the 
self--consistency conditions for 
$\sigma_0$ (and hence for the effective nucleon mass,
$M^*=M+g_\sigma \sigma_0$), as
\begin{mathletters}
\begin{equation}
\frac {d(\Delta \epsilon_\sigma)}{d\sigma_0}
+\frac {\gamma}{(2\pi)^3}g_\sigma \int d \bfm k 
\frac {M^*}{(\bfm k^2+{M^*}^2)^{1/2}}
(\sin^2 \theta_- +\sin ^2 \theta_+)
+\frac {d(\Delta \epsilon_{ren})}{d\sigma_0}=0
\label{selfsig}
\end{equation}
and, for the vector meson field, $\omega_0$, as
\begin{equation}
\omega_0=\frac {g_\omega}{m_\omega^2}\frac {\gamma}{(2\pi)^3}
\int d \bfm k (\sin^2 \theta_- -\sin ^2 \theta_+),
\label{selfomg}
\end{equation}
\label{selfcons}
\end{mathletters}
where $\sin^2 \theta_{\mp}$ are the thermal distribution functions
for the baryons and antibaryons, given through equation (\ref{distr}).

In Eq. (\ref {selfsig}), the first term includes contribution from the
scalar meson condensates. In the mean field approximation for the
scalar mesons in the linear Walecka model, $\lambda_R$=0 and the energy
density from the sigma meson is $\frac {1}{2} m_R^2 \sigma_0^2$. This then
corresponds to the relativistic Hartree approximation, with the
last term in (\ref{selfsig}) being the contribution arising 
from the vacuum polarisation effects from the baryonic sector.

In the next section, we shall consider the meson properties ($\omega$
and $\rho$), calculated from the meson self energy, 
as modified in the medium due to its coupling to nuclear 
excitations.

\section {In-medium masses for $\omega$ and $\rho$ vector mesons}

We now examine the medium modification to the masses of the $\omega$- 
and $\rho$-mesons including the quantum correction effects in the 
hot nuclear matter in the relativistic random phase
approximation. The interaction vertices 
for these mesons with nucleons are given as

\begin{equation}
{\cal L}_{int}=g_{V}\Big (\bar \psi \gamma_\mu \tau^a \psi V_a^{\mu}
-\frac {\kappa_V}{2 M_N} \bar \psi \sigma_{\mu \nu} \tau^a \psi
\partial ^\nu V_a^\mu  \Big)
\label{lint}
\end{equation}
\noindent where $V_a^\mu=\{\omega^\mu,\rho _a^\mu \}$, $M_N$ is the free
nucleon mass, $\psi$ is the nucleon field and $\tau_a=\{ 1, \vec \tau \}$,
$\vec \tau$ being the Pauli matrices. $g_V$ and $\kappa_V$ correspond
to the couplings due to the vector and tensor interactions
to the nucleon fields.  

The vector meson self energy is expressed 
in terms of the nucleon propagator modified by the quantum effects.
This is given as
\begin{equation}
\Pi ^{\mu \nu} (k)=-\gamma_I g_V^2 \frac {i}{(2\pi)^4}\int d^4 p 
Tr \Big [ \Gamma_V^\mu (k) G(k) \Gamma_V^\nu (-k) G(p+k)\Big],
\end{equation}
\noindent where, $\gamma_I$=2, is the isospin degeneracy factor for
nuclear matter. In the above, $\Gamma_V^\mu$ represents the meson-nucleon 
vertex function obtained from (\ref{lint}) and is given by
\begin{equation}
\Gamma_V^\mu (k)=\gamma^\mu \tau_a -\frac {\kappa_V}{ 2 M_N}\sigma^{\mu \nu}
i k_\nu \tau_a
\label{gm}
\end{equation}

The nucleon propagator in the medium is given as
\begin{eqnarray}
G(k) &=& (\gamma ^\mu  {\bar k}_\mu+M_N^*)
\Big [ \frac {1}{\bar k ^2-{M_N^*}^2+i\epsilon}
+2\pi i \delta (\bar k^2 -{M_N^*}^2) \eta (\bar k \cdot u)\big ]\nonumber\\
&=& G_F(k)+G_D(k),
\end{eqnarray}

\noindent where, $\bar k^\mu \equiv (k^0+\Sigma_V^0,\vec k)$ and the 
effective nucleon mass is  $M_N^*=M_N+\Sigma_S$, where the vector and 
scalar self energies are given in terms of the expectation values
for $\omega$ and $\sigma$  fields as $\Sigma_V^0=-g_\omega \omega_0$
and $\Sigma_S=g_\sigma \sigma_0$. In the above, 
$\eta (p\cdot u)=\theta (p\cdot u)f_{FD}(z)+ \theta (-p\cdot u) f_{FD}(-z)$, 
with, $f_{FD}(z)=(1+e^z)^{-1}$, $z=(p \cdot u -\mu ^*)/T$ and $u^\mu$ is
the four-velocity of the thermal bath. The expectation values,
$\sigma_0$ and $\omega_0$, in the presence of quantum 
corrections, are determined self-consistently through the equations
(\ref{selfsig}) and (\ref {selfomg}) for the hot nuclear matter 
\cite{hotnm}. 

The vector meson self energy can be written as the sum of two parts
\begin{equation}
\Pi^{\mu \nu}= \Pi^{\mu \nu}_F+ \Pi^{\mu \nu}_D.
\end{equation}

where,
\begin{mathletters}
\begin{equation}
\Pi ^{\mu \nu}_F=-2i g_V^2 \int \frac {d^4p}{(2\pi)^4}
Tr \Big [ \Gamma_V ^\mu (k) G_F (p) \Gamma_V ^\nu (-k) G_F (p+k)\Big],
\end{equation}
\begin{eqnarray}
\Pi ^{\mu \nu}_D &= & -2i g_V^2 \int \frac {d^4p}{(2\pi)^4}
Tr \Big [ \Gamma_V ^\mu (k) G_F (p) \Gamma_V ^\nu (-k) G_D (p+k)
\nonumber \\
&+& \Gamma_V ^\mu (k) G_D (p) \Gamma_V ^\nu (-k) G_F (p+k)
\nonumber \\
&+& \Gamma_V ^\mu (k) G_D (p) \Gamma_V ^\nu (-k) G_D (p+k)\Big].
\end{eqnarray}
\end{mathletters}
$\Pi^{\mu \nu}_F$ is the contribution arising from the vacuum polarisation
effects, described by the coupling to the $N\bar N$ excitations. 
The shift in the vector meson mass occurs through processes like
$V\rightarrow N\bar N \rightarrow V$, where $N$ represents nucleons
in the medium modified due to the quantum corrections. 
This Feynman part of the self energy, $\Pi ^ {\mu \nu}_F$ is
divergent and needs renormalization. For the $\omega$ meson, 
the tensor coupling is generally small as compared to the vector coupling
to the nucleons \cite {hatsuda1}, and hence is neglected in the present work. 
We use dimensional regularization to separate the divergent parts.
The renormalized vacuum polarization tensor for the $\omega$-meson
is then given as \cite{sourav},
\begin{eqnarray}
\Pi ^{ren}_F (k^2) &=&
\frac {{g_\omega}^2}{\pi^2} k^2 \Big \{ \Gamma (2-n/2)\int _0^1 z (1-z)
\nonumber \\ &-& \int _0 ^1 dz z (1-z) \ln \Big [{M_N^*}^2 -k^2 z (1-z) 
\Big ] \Big \}-\xi,
\end{eqnarray}

in which the last term arises from a counter term added in the
Lagrangian given as
\begin{equation}
{\cal L}_{ct}=-\frac {1}{4} \xi V^{\mu \nu} V_{\mu \nu}.
\end{equation}
The renormalization condition to determine $\xi$ is
\begin {equation}
\Pi ^{ren}_F (k^2)|_{M_N^* \rightarrow M_N}=0.
\end{equation}

We finally arrive at 
\begin{eqnarray}
\Pi_F ^\omega (k^2) &=& \frac {1}{3} Re (\Pi ^{ren} _F)^\mu _\mu
\nonumber \\ & = & -\frac {g_\omega ^2}{\pi^2} k^2
\int _0 ^1 dz z (1-z) \ln \Big [ \frac {{M_N^*}^2-k^2 z (1-z)}{{M_N}^2
-k^2 z (1-z)} \Big ].
\end {eqnarray}

Because of the  tensor interaction, the vacuum self energy for
the $\rho$ meson is not renormalizable. We employ a phenomenological 
subtraction procedure \cite {hatsuda,hatsuda1} to extract the
finite part using the renormalization condition
\begin{equation}
\frac {\partial ^n \Pi _F^\rho (k^2)}{\partial (k^2)^n}|_{M_N^*
\rightarrow M_N}=0,
\end{equation}
with ($n=0,1,2,\cdots \infty$).  Using dimensional regularization
and the above subtraction procedure, we arrive at the following 
expression

\begin{equation}
\Pi ^\rho _F (k^2)=-\frac {g_{\rho}^2}{\pi^2} k^2 
\Big [ I_1 +M_N^* \frac {\kappa_\rho}{2 M_N} I_2 +\frac {1}{2} 
\Big (\frac {\kappa_ \rho}{2 M_N} \Big )^2 (k^2 I_1+{M_N^*}^2 I_2)
\Big]
\end{equation}

where,
\begin{equation}
I_1= \int _0 ^1 dz z (1-z) \ln \Big [ 
\frac {{M_N^*}^2-k^2 z (1-z)}{{M_N}^2 -k^2 z (1-z)} \Big ],
\end{equation}
\begin{equation}
I_2= \int _0 ^1 dz \ln \Big [ 
\frac {{M_N^*}^2-k^2 z (1-z)}{{M_N}^2 -k^2 z (1-z)} \Big ].
\end{equation}

In a hot and dense medium, because of Lorentz invariance and current
conservation, the ground state structure of the polarization tensor
takes the form 
\begin{equation}
\Pi ^{\mu \nu}=\Pi _T (k_0,\vec k) A^{\mu \nu}+
\Pi _L (k_0,\vec k) B^{\mu \nu},
\end{equation}
where the two Lorentz invariant functions $\Pi_T$ and $\Pi_L$, 
characterizing the transverse and longitudinal projection tensors. 
These are obtained through the contractions
\begin{mathletters}
\begin{equation}
\Pi_L=-\frac {k^2}{|\vec k|^2} u^\mu u ^\nu \Pi_{\mu \nu}
\end{equation}
\begin{equation}
\Pi_T=\frac {1}{2} \Big ( \Pi ^\mu _\mu -\Pi_L \Big),
\end{equation}
\end{mathletters}
\noindent with $u_\mu$ being the four velocity of the thermal bath.
In the above, $A^{\mu \nu}$ and $B^{\mu \nu}$ are the transverse
and longitudinal projection operators.
The dispersion relation for the longitudinal (transverse) mode
is obtained as 
\begin{equation}
k_0^2 - |\vec k|^2 -m_V^2 +Re \Pi ^D _{L(T)} (k_0,|\vec k|)
+Re \Pi ^F (k_0,|\vec k|)=0
\label {displt}
\end{equation}

The in-medium mass for the vector meson ($m_V^*$) is defined as the 
lowest zero of equation (\ref {displt}) in the limit 
$\vec k \rightarrow 0$. In this limit,
$\Pi_T^D=\Pi_L^D=\Pi^D$ and we have \cite{sourav},
\begin{equation}
\frac {1}{3} \Pi^\mu _\mu =\Pi =\Pi^D+\Pi^F,
\end{equation}
where the density dependent part for the self energy
is given as
\begin{equation}
\Pi^D (k_0,\vec k \rightarrow 0)
=-\frac {4 g_{V}^2}{\pi^2}\int p^2 dp F(|\vec p|,M_N^*)
\Big [ \sin ^2 \theta_-(\mu ^* ,T)+\sin  ^2 \theta_+(\mu ^* ,T)
\Big ]
\end{equation}
with 
\begin{eqnarray}
F(|\vec p|,M_N^*) &=& \frac {1}{\epsilon^*(p)(4 \epsilon^*(p)^2-k_0^2)}
\Bigg [ \frac {2}{3} (2 |\vec p|^2+3 {M_N^*}^2)+k_0^2 \Big \{ 2 M_N^*
\Big (\frac {\kappa_V}{2 M_N} \Big) \nonumber \\
& +& \frac {2}{3} \Big ( \frac {\kappa_V}{2 M_N}\Big )^2
(|\vec p|^2+3 {M_N^*}^2)\Big \} \Bigg ]
\end{eqnarray}
where $\epsilon ^* (p)=({\vec p} ^2+{M_N^*}^2)^{1/2}$ is the effective
energy for the nucleon. The effective mass of the vector meson is
then obtained by solving the equation
\begin{equation}
k_0^2-m_V^2 +Re \Pi (k_0,\vec k=0) =0.
\label {omgrho}
\end{equation}

As already stated, the $\omega NN $ tensor coupling  is generally small 
(e.g. $\kappa_\omega \simeq 0.1$ in vector dominance model, whereas,
$g_{\omega} \simeq 10$). Hence, we neglect the tensor coupling 
for the $\omega$ meson, in the present calculations. The vector
coupling $g_{\omega}$, along with the scalar coupling $g_{\sigma}$
is fixed from the nuclear matter saturation properties.

For the nucleon-rho couplings, we use the vector and tensor couplings 
as obtained from the N-N forward dispersion relation
\cite{hatsuda1,sourav,grein}. These couplings, however, 
do not take into account 
the medium effects. We also consider the modification of
the $\rho$-meson for the case  when the $\rho$ vector
coupling to the nucleon is determined from the symmetry energy for the 
nuclear matter, thus taking into account the medium dependence
of the coupling. But since the medium dependence of the
tensor coupling is not yet known, we take it as a parameter
and examine the effect of it on the in-medium $\rho$ meson mass.
With the couplings as described above, we
consider the temperature and density dependence
of the $\omega$ and $\rho$ meson masses in the hot nuclear matter
modified due to quantum corrections. 

In the next section, we shall consider effect of quantum corrections
on the strange meson ($\phi$-meson) as coupled to the baryons
through a vector coupling in hyperon-rich matter \cite{shm}. 
The medium modification for $\phi$-meson mass occurs due to 
their coupling to the hyperons, as they do not couple to the nucleons
by OZI rule \cite {hatsuda1}. It is observed that the medium modification
of the $\phi$-meson mass is rather small as compared to the
mass of the vector $\omega$. This is because of the fact the latter has
a significant contribution from the coupling to the nucleon 
sector, and the $\phi$-meson has only contribution from coupling
to hyperons which are rather insenstitive to the medium effects. 

\section {$\phi$- and $\omega$- meson masses modification
in hyperon-rich dense matter}

The interaction for the vector mesons ($\phi$ and $\omega$) with 
the baryons (B=N,$\Lambda$, $\Sigma^{\pm,0}$, $\Xi ^{0,-}$) 
is given through the interaction Lagrangian
\begin{equation}
{\cal L}_{int}=\sum _B g_{\phi B} \bar \psi_B \gamma^\mu 
\psi_B \phi^\mu
\end{equation}

The strange vector meson, $\phi$, couples
to the hyperons, but not to the nucleons \cite {hatsuda1}. Hence 
$\phi$-mesons do not have any mass modification in nuclear matter,
but get modified in strange hadronic matter.
The hyperon-rich dense matter has been studied earlier in a 
nonperturbative treatment \cite{shm} including
the vacuum polarizations from the baryons (nucleons and hyperons)
and scalar meson sector. This has been done in the context of
 structure of neutron stars and so the condition of electrical
charge neutrality has been imposed for the study of the
strange hadronic matter.
In the present section, we study how the vector mesons 
($\omega$ and $\phi$) are modified in such a medium.


The vector meson mass in the medium obtained as a solution
of the dispersion relation (\ref {omgrho}), now gets modified
due to contributions from the hyperons, and is given as, 
\begin{equation}
k_0^2-m_V^2 +\sum _B Re \Pi_B (k_0, \vec k=0) =0.
\label {omgphi}
\end{equation}

To solve the above equation for the in-medium meson mass, we take 
the meson-hyperon couplings as determined through the SU(3)
symmetry \cite{shm,debadesh,pal} given as

\begin{eqnarray}
& &\frac {1}{2}g_{\omega \Lambda}=\frac {1}{2} g_{\omega \Sigma}
=g_{\omega \Xi} =\frac {1}{3} g_{\omega N} \nonumber \\
& & 2 g_{\phi \Lambda}=2 g_{\phi \Sigma}=g_{\phi \Xi}
=-\frac {2 \sqrt 2}{3} g_{\omega N}
\end {eqnarray}

In the following section, we discuss the results obtained in the
present calculations and discuss the effect of quantum corrections
arising from the scalar meson sector over those from the baryonic sector
(corresponding to the relativistic Hartree approximation).

\section {Results and Discussions}

We first discuss the effect of quantum corrections on the 
in-medium nucleonic properties for the hot nuclear matter
\cite{hotnm}. In fig. 1, we plot the temperature dependence
of the nucleon mass for various values of the baryon 
density. The quantum corrections arising from the sigma field
has the effect of the softening of the equation of state,
and, a higher value for the effective nucleon mass as compared 
to the RHA. The effective nucleon mass was seen to increase
with the vacuum polarisation effects arising from the baryons
as compared to the MFT calculations \cite{chin,mishra}. Though 
for $\rho=0$, the effective nucleon mass decreases monotonically
with temperature, for higher
values of densities, the nucleon mass first increases and then 
decreases as a function of temperature \cite{li}. 
The variation in the nucleon mass is very slow upto a temperature, 
T$\simeq$ 150 MeV, beyond which there is a fast decrease. 
Though qualitative features are same for the in-medium mass 
with inclusion of quantum effects from the scalar meson sector, 
it is observed that such effects lead to a higher value for the 
nucleon mass.

We next study the temperature and
density dependence of the vector mesons in the hot nuclear matter.
In figure 2, we plot the $\omega$ meson mass
as a function of the temperature for various densities,
with the $\omega N$ coupling along with the scalar 
coupling $g_\sigma$, determined from the nuclear matter saturation 
properties. In particular, for RHA, 
$C_v^2=g_{\omega}^2 M_N^2/m_\omega^2 =114.7$, 
and with quantum corrections from scalar mesons for self coupling, 
$\lambda_R$=1.8, $C_v^2 =96.45$, 
respectively \cite{mishra}.
In Walecka model, the Dirac sea has been shown to
have a significant contribution over the Fermi sea, leading
to a substantial drop in the omega meson mass in the nuclear matter
\cite{hatsuda,hatsuda1,sourav}. Specifically, at saturation density 
at $T=0$, the decrease in the omega meson mass from the
vacuum value, is around 150 MeV
in RHA, whereas with sigma quantum effects, the drop in the mass
reduces to around 117 MeV. A similar reduction in the mass drop,
as compared to the RHA was also observed when scalar field quantum
corrections were included within an one loop approximation
\cite{serot,pal}, unlike the approximation adopted here, which is
self-consistent and includes multi-loop effects. 
This means that the quantum corrections
do play an important role in the medium modification
of the vector meson masses. In the present work, it is seen that
the quantum corrections from the scalar mesons, over those
arising from the baryonic sector, have the effect of giving rise to
a higher value for the $\omega$-meson mass. This again is related 
to the fact that effective nucleon mass is higher when such quantum
effects are taken into account. 

In figure 3, we illustrate the medium modification for 
the $\rho$ meson mass with the vector and tensor couplings 
to the nucleons being fixed from the NN forward dispersion relation
\cite{hatsuda1,sourav,grein}. The values
for these couplings are given as $g_{\rho N}^2/4\pi$=0.55
and $\kappa_\rho$=6.1. We notice that the decrease in the
$\rho$ meson with increase in temperature is much sharper than
that of the $\omega$ meson. Such a behaviour of $\rho$-meson 
undergoing a much larger medium modification was also observed
earlier \cite {sourav}, with the relativistic Hartree approximation
for the nucleons. This indicates that the tensor coupling plays
a significant role for the $\rho$ meson, which is absent for
the $\omega$ meson. With inclusion of quantum corrections from 
the scalar meson, the qualitative comparison between the  $\omega$
and $\rho$ vector mesons remains the same, though these effects 
are seen to lead to larger values for the vector meson masses.

In the vector and tensor couplings of the $\rho$ meson to the
nucleons as determined from the NN forward scattering
processes \cite {grein} as considered above, any medium dependence 
of these couplings have not been taken into account.
We shall next consider the mass modification
for the $\rho$ meson, with the nucleon $\rho$ coupling,
$g_{\rho}$ as determined from the symmetry energy coefficient,
$a_{sym}$, given as \cite {sym}
$$ a_{sym}=(\frac {g_{\rho}}{m_\rho})^2 \cdot \frac {k_{F0}^3}{12\pi^2}
+\frac {k_{F0}^2}{6(k_{F0}^2+{M_N^*}^2)^{1/2}}$$
where $k_{F0}$ is the Fermi momentum of nuclear matter at saturation
density, $\rho_0$.
In our calculations, we choose $a_{sym}=32.5$ MeV, which gives
the value for the nucleon-rho meson coupling as $g_{\rho}$ = 6.82,
7.0, for RHA and $\lambda_R$ = 1.8, respectively.
However, in considering such a medium dependence of the vector
coupling, it is assumed that the temperature dependence of such
coupling is much smaller compared to its density dependence,
and hence has not been taken into account in the present work.
We take the tensor coupling as a parameter in our calculations.
since the medium dependence of the tensor coupling is not known as yet. 
The in-medium mass of the $\rho$ meson is plotted in figure 4
for baryon density $\rho/\rho_0$=0.5.
It is observed that the $\rho$ meson mass has a strong dependence 
on the tensor coupling. For the same tensor coupling $\simeq$ 6,
the $\rho$ meson mass is much lower ($M_\rho^*/M_\rho \simeq$ 0.33)
when the medium dependence of the vector 
coupling constant is taken into account through the symmetry energy,
as compared to the value ($M_\rho^*/M_\rho \simeq 0.65$), when $g_\rho$
is fixed from the N-N forward scattering processes. 

In figure 5, the effective baryon masses, $M_B^*=M_B+g_\sigma \sigma$,
in the hyperon matter are plotted as a function of the baryon density 
at zero temperature, for RHA, and, with quantum corrections also from 
scalar mesons.
The variation of the hyperon masses are slower as compared to the 
nucleon masses. This is a reflection of a smaller coupling of
hyeprons to the scalar sigma field \cite{shm,debadesh}.
Further, we note that the decrease in the effective baryon mass
with density, is slower when quantum corrections from the sigma
mesons through condensates are taken into account as was the case
for nuclear matter \cite{mishra}.

The medium modification for the vector mesons ($\omega$ and $\phi$) 
in the hyperonic matter are plotted in figure 6. 
One observes that the strange meson $\phi$ has smaller modification 
to the mass as compared to that of the $\omega$ meson. 
This is due to the fact that $\phi$-meson does not couple 
to the nucleons and also, the hyperon masses are rather insensitive 
to the changes in the baryon densities as may be seen in figure 5.
The strange meson ($\phi$) mass modification observed  as small
compared to the $\omega$ and $\rho$ meson masses is in line
with the earlier observations \cite{hat,hatsuda1,pal}.

\section{Summary}

To summarize, we have considered in the present paper, the vector
meson mass modifications due to quantum correction effects in the hot 
and dense matter. The baryonic properties as modified due to
such effects subsequently determine the vector meson
masses in the hot and dense hadronic matter. It has been recently
emphasized that the Dirac sea contribution dominates over the
fermi sea, thus implying that the vacuum polarisation effects arising
from the baryonic sector do play an important role in the vector meson
properties in a medium. In the present work, we study the effect
of quantum corrections arising from the scalar meson over those
resulting from the baryon sector (given through
the summing over baryonic tadpole diagrams in the relativistic
Hartree approximation). The quantum corrections from $\sigma$-meson
give rise to a higher value for the vector meson masses. Also,
it is observed that the strange vector meson ($\phi$) has smaller
medium modification in the hyperon matter as compared to the
$\omega$-meson similar to earlier observations. This is a reflection 
of the facts that, firstly, the hyperon masses have much smaller
 modification as compared to that of nucleons and, also, secondly, 
the $\phi$-meson does not couple to the nucleons, whereas
$\omega$-meson does have appreciable contribution from the coupling
to the nucleons. The vector meson properties being modified in
a medium have an important role to play in the dilepton production
in relativistic heavy ion collisions and it will be interesting to
investigate how the dilepton yield gets affected due to the
quantum corrections in the hot and dense matter. This and 
related problems are under investigation.

\begin{figure}
\psfig{file=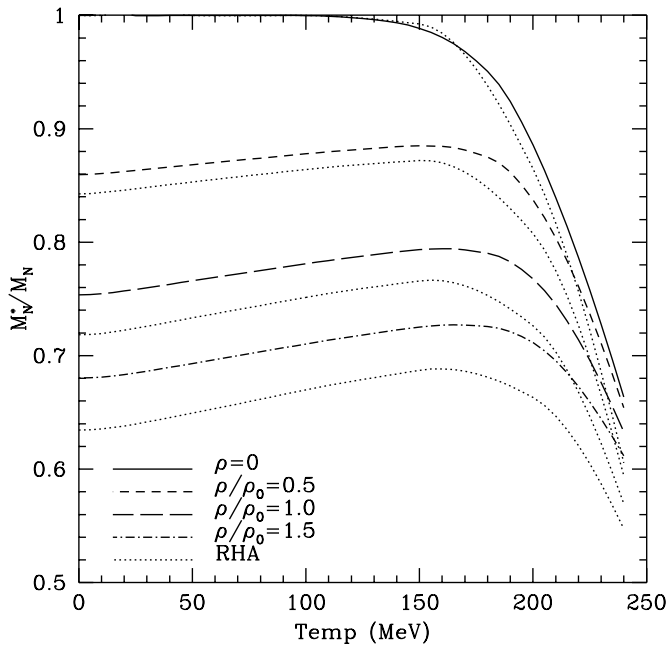,width=9cm,height=9cm}
\caption{Effective nucleon mass as a function of the temperature
for various values of the baryon densities for RHA and $\lambda_R$=1.8.
The quantum corrections from the scalar meson sector leads to
an increase in the in-medium nucleon mass.}
\end{figure}
\begin{figure}
\psfig{file=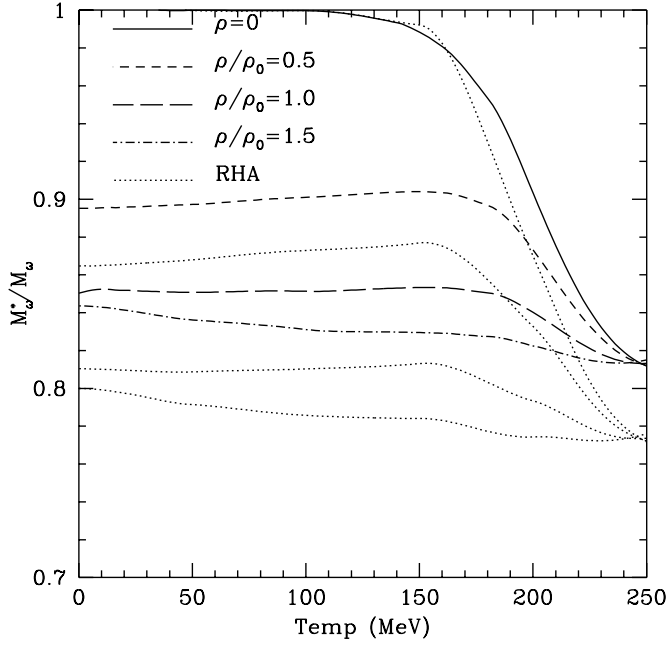,width=9cm,height=9cm}
\caption{In-medium $\omega$-meson mass as a function of temperature.
The sigma meson quantum effects lead to an increase in the
vector meson mass.} 
\end{figure}
\begin{figure}
\psfig{file=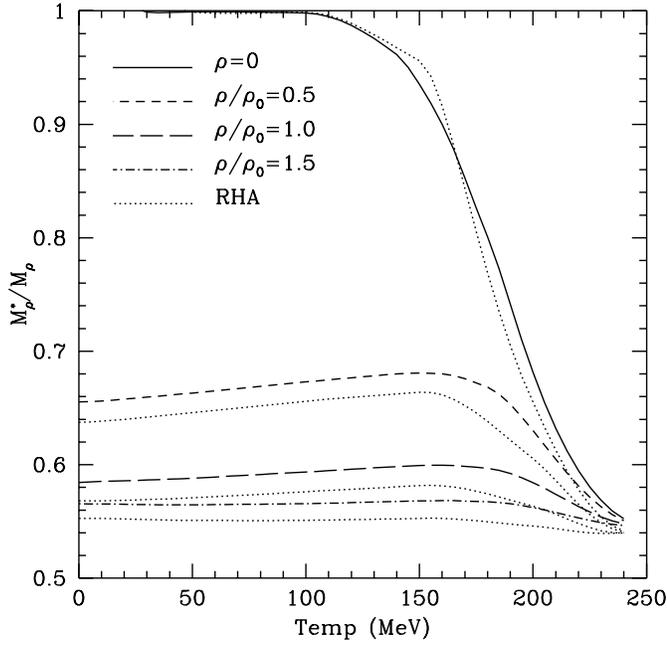,width=9cm,height=9cm}
\caption{In-medium $\rho$-meson mass as a function of temperature
and density for $g_{\rho}^2/4\pi=0.55$, $\kappa_\rho$=6.1.
The decrease is sharper than that of the $\omega$ meson.
The quantum corrections from scalar meson lead to a higher
meson mass.} 
\end{figure}
\begin{figure}
\psfig{file=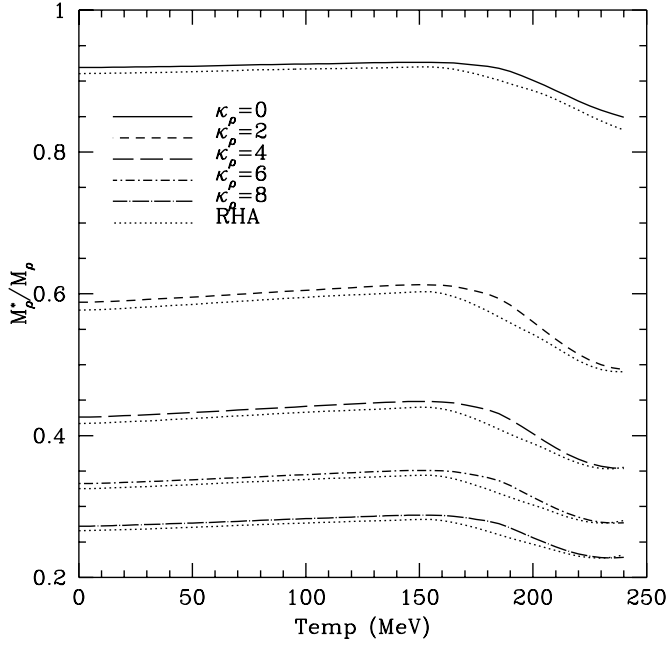,width=9cm,height=9cm}
\caption{In-medium $\rho$-meson mass as a function of temperature
and density, with the vector coupling $g_{\rho}$ fitted from the symmetry
energy for the nuclear matter, and for various values of the tensor 
coupling. The quantum corrections from scalar meson increase
the vector meson mass.} 
\end{figure}
\begin{figure}
\psfig{file=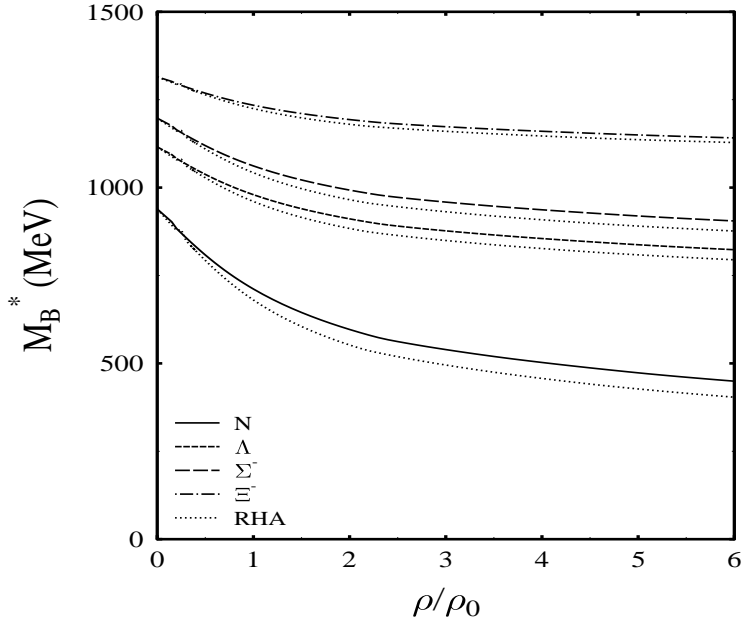,width=11cm,height=11cm}
\caption{Effective baryonic masses as a function of 
density for hyperonic matter. The nucleon masses vary much faster
than the hyperon masses. The quantum effects lead to an
increase in the effective masses as compared to RHA.}
\end{figure}
\begin{figure}
\psfig{file=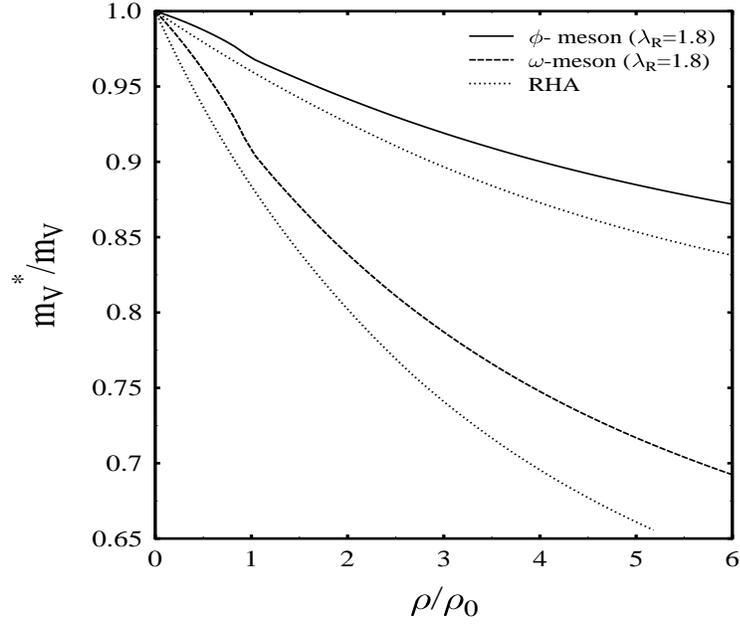,width=11cm,height=11cm}
\caption{Effective vector meson masses ($\phi$ and $\omega$)
plotted as a function of density for hyperon matter
for RHA and $\lambda_R$=1.8. The variation in the strange meson mass
is rather slow as compared to the $\omega$ meson.} 
\end{figure}
\end{document}